\documentclass{amsproc}
\usepackage{amsmath,amscd}

\usepackage{epsfig}

\usepackage{amssymb}

\usepackage{amsmath}

\usepackage{latexsym}

\newtheorem{prop}{Proposition}[section]

\theoremstyle{remark}

\numberwithin{equation}{section}

\newcommand{\mbz}{{\mathbb Z}}

\newcommand{\mbr}{{\mathbb R}}

\newcommand{\mpx}{{\mathcal P}X}

\def\genf#1#2{\buildrel#2\over{\mathfrak#1}}

\begin{document}

\title{Negative Forms and Path Space  Forms}


\author{Saikat Chatterjee} \author{Amitabha Lahiri}
\address{{\rm Saikat Chatterjee and Amitabha Lahiri}\\
S.~N.~Bose National Centre for Basic Sciences \\ Block JD,
  Sector III, Salt Lake, Kolkata 700098 \\
  West Bengal, INDIA}
\email{saikat@bose.res.in, amitabha@bose.res.in}

\author{Ambar N. Sengupta}
\address{{\rm Ambar Sengupta}\\ Department of Mathematics,
  Louisiana State University\\  Baton 
Rouge, Louisiana 70803\\ USA}
\email{sengupta@math.lsu.edu}
\urladdr{http://www.math.lsu.edu/\symbol{126}sengupta}




\date{November 30, 2007.}


\begin{abstract} 
We present an account of negative differential forms within a
natural algebraic framework of differential graded algebras, and
explain their relationship with forms on path spaces.

\end{abstract}

\maketitle

\section{Introduction}\label{intro}

In this paper we \begin{itemize} \item[(i)] demonstrate that
negative-degree differential forms, which have been used in the
physics literature, can be formalized mathematically by means of a
Koszul complex;
\item[(ii)] derive certain fundamental properties of a natural map
  ($I$
in our (\ref{E:mapOmegaps}), involving the Chen integral,  from the
space of generalized differential forms to   differential forms over
path space.
\end{itemize}

Differential forms of negative degree have been used in a number of
works~\cite{NurRob2001,NurRob2002,CLG}. Such forms were introduced
by Sparling \cite{Spar} in the context of twistors. In
\cite{NurRob2001} these forms were used to cast Einstein's equations
for gravity in a convenient form. The algebra and calculus of these
forms and related geometrical objects were developed further
in~\cite{NurRob2002,CLG}.

Another motivation for studying differential forms of negative
degree comes from the fact that the flat connections of a gauge
theory made out of such generalized form connections are very
similar to the flat connections found in the study of integrable
systems~\cite{Alvarez:1997ma} and four dimensional $B\wedge F$
theories~\cite{Cat}. The particular type of integrals over paths
(given in our map $I$ in \ref{E:mapOmegaps}) appears naturally in
formulas for the variation of parallel transport along
paths~\cite{Sengupta}.  Even more remarkably, they also appear to
closely resemble flat connections on a gerbe~\cite{Breen:2001ie}.
While we will not attempt to relate these diverse contexts, nor
explain the similarities among the structures, we believe that it
may be worthwhile to look at these objects from different
perspectives.

In the present paper we describe a coherent, unifying algebraic
framework for the calculus of such forms and relate them to forms on
path spaces. More specifically, in section \ref{S:Construct} we
shall show how to realize negative differential forms, as well as
the usual ones, as elements of a Koszul complex (we will explain the
terminology). Then, in section \ref{S:Path}, we show how the first
order Chen integral helps set up a natural mapping from generalized
differential forms to forms on the space of paths, and establish
some of its essential properties.

For ease of reference we have included an appendix containing the
relevant essentials of differential graded algebras. Some of the
notions we have used in this paper could be recast in the language
of supermanifolds, but we have not done so.

\section{Construction of the Complex of Generalized Forms}
\label{S:Construct}

A generalized $p$-form, for any integer $p\geq -1$, has been
defined (see~\cite{NurRob2001}) as an ordered
pair $ (a_p,a_{p+1})$ of ordinary differential forms, where $a_p$
is a $p$-form, equal to $0$ in case $p$ is $-1$.  The product of
two such forms is defined to be
\begin{equation}\label{E:defgeneralwedge1}
(a_p,a_{p+1})\wedge (b_q,b_{q+1})=(a_p\wedge b_q, a_p\wedge
  b_{q+1}+(-1)^qa_{p+1}\wedge b_q)
\end{equation}
A differential calculus of such forms has been developed in the
cited works, especially in \cite{CLG}. For example, the exterior
differential is defined by
\begin{equation}\label{E:defdgeneral}
d(a_p,a_{p+1})=(da_p+(-1)^{p+1}ka_{p+1}, da_{p+1})
\end{equation}
where $k\in\mbr$ is a `parameter' in this system (in
principle, we could let $k$ be an abstract variable).

Our objective in this section is to present a new construction of
the generalized forms showing how they fit in naturally in the
category of differential graded algebras. The definitions
(\ref{E:defgeneralwedge1}) and (\ref{E:defdgeneral}) will then
appear in a completely natural way.

\subsection{\sl The standard algebra of differential forms}

Consider the graded algebra of differential forms over an open
subset $U$ of $\mbr^N$:
 $$\Omega\stackrel{\rm def}{=}\Omega(U)\stackrel{\rm def}{=}
 \bigoplus_{p\in{\mbz}}\Omega_p,$$ where $$\Omega_p=\Omega_p(U)$$
 is the space of all smooth differential forms over $U$ of degree
 $p$, with
 $$\Omega_p=\{0\}\quad\hbox{for all $p<0$}$$
 On $\Omega$ there is the exterior differential
 $$d:\Omega\to \Omega$$
 which is a superderivation of degree $1$, and satisfies
 $$d^2=0.$$
 Thus, $(\Omega,d)$ is a complex, and is the basic example of
  a   differential
 graded algebra. (See the Appendix for relevant concepts about
 differential graded algebras.)

\subsection{\sl An auxilliary complex}

 Consider next a new graded algebra
 $$\Omega'=\bigoplus_{p\in {\bf Z}}\Omega'_p,$$
 where $\Omega'_p$ is $\{0\}$ for all $p$ except
\begin{equation}\label{def:Omegaprime}
\Omega'_0={\mbr},\qquad \Omega'_{-1}={\mbr}\zeta,
\end{equation}
where $\zeta$ is an abstract symbol. Multiplication is defined in
the natural way, with the understanding that
$$\zeta^2=0.$$

We can define a superderivation $d $ on $\Omega'$ of degree
$1$. This has to be zero on all elements of $\Omega'_0$ because
$\Omega'_1=\{0\}$, and so is uniquely specified by its value on
$\zeta$. We denote this value by $k$:
$$d\zeta=k\in\mbr.$$

\subsection{\sl The tensor product complex}

Now we introduce the tensor product complex:
$${\hat\Omega}(U)=\Omega(U)\otimes\Omega'$$
as a graded algebra. The degree $p$ subspace of the tensor product
of two graded algebras $A$ and $B$ is, by definition,
$$ (A\otimes B)_p=\bigoplus_{j\in{\mathbb Z}}A_j\otimes B_{p-j}$$
Thus, in our case,
$${\hat\Omega}_p=(\Omega_p\otimes \Omega'_0)\oplus
(\Omega_{p+1}\otimes\Omega'_{-1})\simeq
\Omega_p\oplus (\Omega_{p+1}\otimes {\mbr}\zeta)$$
and a typical element of ${\hat\Omega}_p$ is  of the form
\begin{equation}
\label{E:alphap}
\alpha_p=a_p+a_{p+1}\zeta,
\end{equation}
where $a_p\in\Omega_p$, $a_{p+1}\in\Omega_{p+1}$, and
we have identified $a\otimes 1$ with $a$, and $1\otimes \zeta$ with
$\zeta$. To compare our development of generalized forms with
that in earlier works one should use the identification
\begin{equation}
\label{E:identif}
a_p+a_{p+1}\zeta \leftrightarrow (a_p,a_{p+1})
\end{equation}
Multiplication of homogeneous elements is given explicitly by
\begin{eqnarray*}
(a_p+a_{p+1}\zeta)(b_q+b_{q+1}\zeta)&=&
a_pb_q+a_pb_{q+1}\zeta+a_{p+1}{\zeta}b_q+a_{p+1}{\zeta}b_{q+1}\zeta
\\
&=& a_pb_q+\bigl(a_pb_{q+1}+(-1)^qa_{p+1}b_q\bigr)\zeta
\end{eqnarray*}
which `explains' the multiplication of generalized forms defined
earlier in (\ref{E:defgeneralwedge1}). Here, as is standard in
super-algebra, we have inserted $(-1)^{rs}$ when moving an element
of degree $r$ past an element of degree $s$ (see
(\ref{gradedproduct}) in the Appendix).

As shown in the Appendix, the two operators $d$ on $\Omega(U)$ and
$\Omega'$ combine to yield a degree $+1$ superderivation $d$ on
${\hat\Omega}$ also satisfying
$$d^2=0$$ The action of $d$ is expressed by
\begin{equation}
\label{E:dalpha}
d\alpha_p=da_p
+(da_{p+1})\zeta+(-1)^{p+1}a_{p+1}d\zeta=da_p+(-1)^{p+1}ka_{p+1}
+(da_{p+1})\zeta
\end{equation}
where $\alpha_p$ and other notation are as in (\ref{E:alphap}).
This `explains' the earlier {\em ad hoc} definition of the exterior
differential in (\ref{E:defdgeneral}). It may also be noted that we
have here a cochain analog of the notion of a {\em mapping cone}
for the chain map: $a_p\mapsto k(-1)^pa_p$ (see, for instance,
section 4.3 in~\cite{Spanier}).

For a smooth manifold $M$, each point has a neighborhood $U$
diffeomorphic, by a chart, to an open subset of Euclidean space,
and so ${\hat\Omega}(U)$ makes sense. These algebras fit together
consistently, just as in the usual calculus of differential forms,
to yield differential forms over the whole manifold $M$.

Thus, the tensor product of differential graded algebras provides
the correct structure for the generalized forms.    We will not
develop this structure further here to include generalized vector
fields (see~\cite{CLG} for the latter).

\subsection{\sl The tensor product with the Koszul complex}

We will now present a natural algebraic framework for a broader
class of generalized differential forms introduced by Sparling
\cite{Spar} in the context of twistor theory.

Let $V$ be a finite dimensional vector space, and $\bigwedge V$ the
exterior algebra over $V$:
\begin{equation}
\label{E:ExtV}
\bigwedge V=\bigoplus_{p\in {\mbz}}{\bigwedge}{}^{p}V
\end{equation}
where, of course, the negative exterior powers are, by definition,
$0$. We consider this same algebra with the grading inverted:
\begin{equation}\label{E:ExtVneg}
(\bigwedge V)^{-1}=\bigoplus_{p\in {\mbz}}{\bigwedge}{}^{-p}V
\end{equation}
so that
$$(\bigwedge V)^{-1}_p={\bigwedge}{}^{-p}V$$
Fix a basis $\zeta_1,...,\zeta_n$ of $V$, and elements
$k_1,...,k_n\in\mbr$. Requiring that
$$d\zeta_i=k_i\quad\hbox{for each $i=1,...,p$}$$
specifies a unique degree $+1$ superderivation
$$d:(\bigwedge V)^{-1}\to (\bigwedge V)^{-1}$$
Then
$$d^2=0$$
and the complex thus obtained is essentially what is known as a
Koszul complex (see Section 16.10 in~\cite{Lang}).

 Now form the graded tensor product algebra
\begin{equation}\label{E:hatOmega}
{\hat\Omega}=\Omega\otimes (\bigwedge V)^{-1}
\end{equation}
As explained in the Appendix (Proposition A.1), the two differential
operators $d$ combine to yield a superderivation of degree $+1$ on
${\hat\Omega}$ with $d^2=0$.

The typical element of ${\hat\Omega}_p$ has now the form
\begin{equation}
\alpha_p= a_p + \sum_{i=1}^n a_{p+1}^i\zeta_i + \sum_{i_1,i_2=1}^n
a_{p+2}^{i_1i_2}\zeta_{i_1}\zeta_{i_2} + \cdots +
a_{p+n}\zeta_1...\zeta_n
\end{equation}
These are the generalized forms of~\cite{Robinson2003},
and the differential structure is also the same.

Note that in the preceding construction, we could also take
$k_1,...,k_n$ to be abstract variables (indeterminates) and produce
forms with coefficients in the algebra $\mbr[k_1,...,k_n]$ or
quotients thereof.

\section{Path Space Forms}\label{S:Path}
\setcounter{equation}{10}

Forms on path spaces have been studied and developed from many
points of view by numerous authors (\cite{Cat}, to cite just one).
For analytic purposes the natural path space on a manifold $X$ is
the space of $H^1$ maps of an interval into
$X$. Driver~\cite{Driver} has developed the machinery for this
setting, showing in particular that the space of such paths forms a
Hilbert manifold. Driver also considered forms of a certain type on
path space.  For more algebraic purposes, Chen \cite{KTChen1,
KTChen} developed a framework of differentiable spaces which is
most convenient for the study of iterated integrals of differential
forms. (The notion of differentiable spaces has been developed much
further in its incarnation as diffeological spaces;
see~\cite{Igles}). We shall look at these approaches very briefly,
before focusing on our setting of smooth paths. We shall limit
ourselves to the minimal essentials here in order to demonstrate
the relationship between path-space forms and generalized
differential forms.

\subsection{\sl Path Spaces}

Let $X$ be a fixed smooth manifold. For the space ${\mathcal
P}_{1}X$ of $H^1$ paths $\gamma:[0,1]\to X$ (i.e. those which, in
local charts, are absolutely continuous with almost-everywhere
defined square-integrable derivatives), the tangent space
$T_{\gamma}{\mathcal P}_1X$ consists of $H^1$ vector fields along
$\gamma$. A differential form on this path space would then be a
skew-symmetric multilinear continuous linear functional on powers
of these tangent spaces.

 A piecewise smooth path in $X$ is a continuous mapping $c:[0,1]\to
X$ for which there exists a partition $0=t_0<t_1<\ldots<t_n=1$ such
that $c|[t_{i-1},t_i]$ is $C^{\infty}$ for $i\in\{1,...,n\}$. The
set of all piecewise smooth paths in $X$ will be denoted ${\mathcal
P}X$, and will be equipped with the compact-open topology. The
latter is to ensure that a map $\phi:Y\to {\mathcal P}X$ is
continuous if and only if the corresponding map
$${\tilde\phi}:  [0,1]\times Y\to X:(t,y)\mapsto \phi(t)(y)$$  is
continuous.

A {\em plot} on $\mpx$ (as defined in~\cite{KTChen}) is a
continuous map $\phi:U\to \mpx$, with $U$ a closed convex region in
some Euclidean space $\mbr^m$, such that there is a partition
$0=t_0<t_1<\ldots<t_n=1$ for which the restrictions
$${\tilde\phi}|  [t_{i-1},t_i]\times U \to X$$
are all $C^{\infty}$.

\subsection{\sl Differential forms on path spaces}

A differential form ${w}$ of degree $p$ on $\mpx$ is specified
by a smooth $p$-form ${w}_\phi$ on $U$ for each plot $\phi:U\to
\mpx$ with the consistency condition that
$$f^*{w}_\phi={w}_{\phi\circ f}$$
for any $C^{\infty}$ map $f$ into the domain of any plot
$\phi$. Let $\Omega_p(\mpx)$ be the set of all $p$-forms on $\mpx$;
note that $\Omega_0(\mpx)$ consists of smooth functions
on $\mpx$, and $\Omega_p(\mpx)$ is, by definition, $0$ for
$p<0$. Then
$$\Omega(\mpx)=\bigoplus_{p\in\mbz}\Omega_p(\mpx)$$ is a graded
supercommutative algebra, as usual, and there is a natural exterior
differential operator $d$, a superderivation of degree $+1$, on
$\Omega(\mpx)$ specified through
\begin{equation}\label{E:domegaphi}
(d{w})_\phi=d({w}_\phi)
\end{equation}

Naturally,   pullbacks of path-forms  are defined (see~\cite{KTChen}
for details) in such a way that
$${w}_\phi=\phi^*{w}.$$

The analytic approach suggests that a convenient example of a
$p$-form on $\mpx$ should be given as follows. Suppose
$\phi:U\to\mpx$ is a plot with $\phi(u)=\gamma$ for some point
$u\in U$, and $V_1,...,V_p\in T_uU$ are vectors. Then we have $p$
vector fields $\phi_*V_i$ along $\gamma$ given, at all but finitely
many values of $t$, by
$$\phi_*V_i(t)=D_{V_i}{\tilde\phi}(t,\cdot), $$
where on the right we have the derivative of the map
$${\tilde\phi}(t,\cdot):U\to X:y\mapsto {\tilde\phi}(t,y)$$
at $u$ along $V_i\in T_uU$.

So if ${w}$ is a $p$-form along $\gamma$ (a section of the
pullback of $\wedge^pT^*X$ over $\gamma$) then
$${w}_\phi:(V_1,...,V_p)\mapsto
\int_0^1{w}(\phi_*V_1,...,\phi_*V_p)\,dt$$
provides, at least formally, a $p$-form on $\mpx$ in the sense of
Chen explained above.

\subsection{\sl From   forms on spaces to forms on path spaces}

Consider a $(p+1)$-form ${w}$ on $X$, where $p$ is an integer $\geq
-1$. Consider a plot $\phi:U\to \mpx$, which corresponds, as usual,
to a map ${\tilde\phi}:[0,1]\times U \to X$.
Then ${\tilde\phi}^*{w}$ is a $(p+1)$-form on $[0,1]\times U$ which
we may write as
\begin{equation}\label{E:phiomega}
{\tilde\phi}^*{w}=dt\wedge
{\dot{w}}_{\phi}(t)+{w}_{\phi}(t)
\end{equation}
where ${\dot{w}}_{\phi}(t)\in\Omega_p(U)$ and
${w}_{\phi}(t)\in\Omega_{p+1}(U)$, both pulled back to forms over
$[0,1]\times U $ via the projection on the second factor. Define the
first order Chen integral $\int_0^1{w}$ to be the $p$-form on
$\mpx$ specified through
\begin{equation}\label{E:defintIomega}
\left(\int_0^1{w}\right)_{\phi}\stackrel{\rm
  def}{=}\int_0^1{\dot{w}}_{\phi}(t)\,dt\in
\Omega_p(U)
\end{equation}
Note that this (first-order) {\em Chen integral carries a
  $(p+1)$-form on $X$ to a $p$-form on} $\mpx$.  A zero-form,
  i.e. a function, is taken to $0$ because the corresponding
${\dot{w}}_{\phi}(t)$ is $0$.  Thus we can take the Chen  integral
$$\int_0^1:\Omega(X)\to\Omega(\mpx)$$ to be a degree $-1$ linear
map (hence equal to $0$ in all negative degrees).

Let
\begin{equation}
\label{E:defevt}
{\rm ev}_t:\mpx\to X:\gamma\mapsto\gamma(t)
\end{equation}
be the evaluation map.  Then, (\ref{E:phiomega}) implies that
\begin{equation}
\label{E:evstaromega}
({\rm ev}_t^*{w})_{\phi}=({\rm
  ev}_t\circ\phi)^*{w}={w}_{\phi}(t),
\end{equation}
where ${w}_{\phi}(t)$ is as defined through
(\ref{E:phiomega}). Thus we have the degree-preserving map
\begin{equation}\label{E:evtstar}
{\rm ev}_t^*:\Omega(X)\to {\Omega}(\mpx)\end{equation}
which is, of course, $0$ on all $\Omega_p$ for $p<0$ (these
subspaces being all $0$).

The following observation is useful for us:
\begin{prop}\label{P:stokes} If ${w}\in\Omega(X)$ then
\begin{equation}\label{E:ChenHomotopy}
\int_0^1 d{w}+d\int_0^1{w} = {\rm
  ev}_1^*{w} - {\rm ev}_0^*{w}\end{equation}
Thus, the first Chen integral $\int_0^1$ provides a chain homotopy
between the maps   $ {\rm ev}_1^*$ and $ {\rm ev}_0^*$.
\end{prop}

\underline{Proof} If $w\in \Omega_p(X)$ with $p<0$ then $w$ and all
terms in (\ref{E:ChenHomotopy}) are $0$.  So we assume that $w$ is
a form of non-negative degree. Applying $d$ to (\ref{E:phiomega})
we obtain
\begin{equation}
\label{E:dphiomega}
{\tilde\phi}^*d{w}=-dt\wedge d_U{\dot{w}}_{\phi}(t) +
dt\wedge\frac{\partial{w}_{\phi}(t)} {\partial t} +
d_U{w}_{\phi}(t)
\end{equation}
where we have used the subscript in $d_U$ to stress that it is the
differential of a form on $U$ (lifted back to $[0,1]\times U$). So
\begin{eqnarray*}
\left(\int_0^1d{w}\right)_\phi&=&-\int_0^1
\bigl(d_U{\dot{w}}_{\phi}(t)\bigr)\,
dt + \int_0^1\frac{\partial{w}_{\phi}(t)}{\partial
  t}\,dt \\
&=&-d\left(\int_0^1{w}\right)_{\phi}+{w}_{\phi}(1) -
{w}_{\phi}(0)
\end{eqnarray*}
On using (\ref{E:domegaphi}), the first term on the right is
$-(d\int_0^1{w})_\phi$. The other terms can be read off using
(\ref{E:evstaromega}), showing that
\begin{equation}
\left(\int_0^1d{w}\right)_\phi=-\left(d\int_0^1{w}\right)_\phi
+\bigl({\rm ev}_1^*{w}-{\rm  ev}_0^*{w}\bigr)_\phi.
\end{equation}
This yields the desired formula (\ref{E:ChenHomotopy}). \fbox{QED}

Let us note here that formulae of this type have been useful in a
number of contexts (Proposition 4.1.2 in~\cite{KTChen}, and
Theorem 6.14.1 in~\cite{BottTu}, for instance).

\subsection{\sl Generalized forms over $X$ and forms on $\mpx$}

We can now specify a mapping from the elements of ${\hat\Omega}_p$,
for $X$, to $\Omega_p(\mpx)$, i.e. from generalized $p$-forms over
$X$ to $p$-forms over the path space $\mpx$:
\begin{equation}
\label{E:mapOmegaps}
{w}_p+{w}_{p+1}\zeta\mapsto {\rm ev}_1^*{w}_p-{\rm
  ev}_0^*{w}_p+k(-1)^{p+1}\int_0^1{w}_{p+1}
\end{equation}
where ${w}_p\in\Omega_p(X)$ and ${w}_{p+1}\in\Omega_{p+1}(X)$;
when $p<0$, each term on the right  is $0$ by definition.
These maps combine to yield a degree-preserving linear map
\begin{equation}\label{E:Icohom}
I:{\hat\Omega}(X)\to \Omega(\mpx).\end{equation}

Our aim is to explore the mathematical structure of the map $I$. As
to its origins, let us note only in brief that, for instance, the
`Duhamel formula', known sometimes also as a non-abelian Stokes
theorem, which gives the variation of parallel-transport along a
path (see for example Proposition 3.6 in~\cite{Sengupta}) can be
expressed in the form of the right side in (\ref{E:mapOmegaps}).

Now we can prove:
\begin{prop}\label{P:dI} The mapping $I$ commutes with the
differential $d$:
$$Id=dI$$

\end{prop}
\underline{Proof}.  For any ${w}_p\in\Omega_p(X)$ and
${w}_{p+1}\in\Omega_{p+1}(X)$, where $p\in\mbz$, we have
\begin{eqnarray*}
dI({w}_p+{w}_{p+1}\zeta) &=& d\left({\rm ev}_1^*{w}_p-{\rm
ev}_0^*{w}_p+k(-1)^{p+1}\int_0^1{w}_{p+1}\right)\\
&=& {\rm ev}_1^*d{w}_p-{\rm ev}_0^*d{w}_p +
k(-1)^{p+1}d\int_0^1{w}_{p+1}\\
&=&  {\rm ev}_1^*d{w}_p-{\rm ev}_0^*d{w}_p\\
&&\qquad +k(-1)^{p+1}\left[-\int_0^1d{w}_{p+1}+{\rm
    ev}_1^*{w}_{p+1}-{\rm ev}_0^*{w}_{p+1}\right]\\
&=& {\rm ev}_1^*\bigl(d{w}_p+k(-1)^{p+1}{w}_{p+1}\bigr)-{\rm
  ev}_0^*\bigl(d{w}_p+k(-1)^{p+1}{w}_{p+1}\bigr)\\
&&\qquad\qquad +k(-1)^{p+1+1}\int_0^1d{w}_{p+1}\\
&=& I\bigl(d{w}_p+k(-1)^{p+1}{w}_{p+1} +
(d{w}_{p+1})\zeta\bigr)\\
&=& Id({w}_p+{w}_{p+1}\zeta).\hskip 2in \fbox{QED}
\end{eqnarray*} %

We turn to further properties of  $I$ (we make the natural
assumption that $k\neq 0$).

\begin{prop}\label{P:Iinjec} For the map $I$ we have:
\begin{itemize}
\item[(i)] the kernel $\ker I$ of the linear map $I$ is contained
  inside ${\hat\Omega}_{-1}\oplus {\hat\Omega}_0$ and
   consists of all elements of the form
  $g\zeta+d(f\zeta)$ with $f,g$ running over $\Omega_0(X)$;
\item[(ii)] $I$ is injective on $\bigoplus_{p\geq 1}{\hat\Omega}_p$;
\item[(iii)] for $\alpha\in{\hat\Omega}_p$, with $p\geq 0$,
  $dI(\alpha)$ is $0$ if and only if $d\alpha$ is
  $0$. \end{itemize}
\end{prop}

\underline{Proof}. Note that (i) implies (ii), and then (iii)
follows  from the fact that $Id=dI$. Thus, it will suffice to prove
(i).

Consider $I$ on ${\hat\Omega}_p$ for $p\geq 1$. Suppose
  ${w}_p\in \Omega_p(X)\,,{w}_{p+1} \in
\Omega_{p+1}(X)\,,$ are such that
$$I({w}_p+{w}_{p+1}\zeta)=0$$
 Contracting both sides on $p$ arbitrary vector fields $Q_1,
 \ldots, Q_p \in T_{\gamma}(\mpx)$ we have
\begin{eqnarray}
{w}_p(Q_1(t), Q_2(t), \cdots, Q_p(t)) \mid_{\gamma(1)} -
  {w}_p(Q_1(t), Q_2(t), \cdots, Q_p(t)) \mid_{\gamma(0)} &&
  \nonumber     \\
+(-1)^{p+1}k\int_\gamma{w}_{p+1} (\dot\gamma(t), Q_1(t), Q_2(t),
  \cdots,
  Q_p(t))=0, && \label{E:oemgapQ}
\end{eqnarray}
  By choosing a set of vector fields $Q_i$, with any specified
  (arbitrary) values at $0$, and equal to zero everywhere except
  near a small neighborhood of $0$, we see that the form ${w}_p$
  must be $0$. Then, proceeding with ${w}_{p+1}$, and (essentially)
  taking the derivative with respect to $t$ at $t=1$, we see that
  $\iota_{{\dot\gamma}(t)}{w}_{p+1}$ is $0$ on any vectors
  $Q_1(1),\cdots, Q_p(1)\in T_{\gamma(1)}X$. The path $\gamma$
  being arbitrary, it follows then that ${w}_{p+1}$ is, in fact,
  also $0$.

 Next we turn to the case $p=0$. We write $f$ in place of
 ${w}_0$. Equation (\ref{E:oemgapQ}) still holds, except no vectors
 $Q_i$ are involved:
 \begin{equation}\label{E:dfomega1}
 f\bigl(\gamma(1)\bigr) - f\bigl(\gamma(0)\bigr) -
 k\int_0^1{w}_1\bigl({\dot\gamma}(t)\bigr)\,dt = 0
\end{equation}
 Differentiating this at the `upper' end point shows that, since
 $\gamma$ is arbitrary,
 $${w}_1=k^{-1}df$$
 (assuming $k\neq 0$). Thus, the kernel of $I$ consists of all
 generalized forms of the type
 $$f+k^{-1}(df)\zeta= d(k^{-1}f\zeta),$$
where $f$ runs over all smooth functions over $X$. We can then
absorb the factor $k^{-1}$ into $f$. Lastly, note that $I$ is $0$
on all elements $g\zeta\in{\hat\Omega}_{-1}$, by
definition. \fbox{QED}

Using the map $I$ we can transfer the wedge product from
$\hat\Omega$ onto the image $I(\hat\Omega(X))$ in $\Omega(\mpx)$,
by requiring that the following diagram commute:
\begin{equation}
\label{CD:generalwedge1}
\begin{CD}
\hat\Omega(X) @> I >> I(\hat\Omega(X))\subset\Omega(\mpx) \\
@V\wedge VV  @VV\wedge' V \\
\hat\Omega(X) @> I >> I(\hat\Omega(X))\subset \Omega(\mpx)
\end{CD}
\end{equation}
Because $I$ is not injective this has to be viewed  as operating
only on forms of degree $\geq 1$.

The wedge product on the left vertical arrow is given by
Eq.~(\ref{E:defgeneralwedge1}). For any $a_p, a_{p+1},
b_q, b_{q+1} \in \Omega(X)\,,$ with $p\geq 1$, we have
\begin{eqnarray}
\genf a p &\stackrel{\rm def}{=}& {  I}(a_p + a_{p+1}\zeta) =
 {\rm ev}_{1}^{*}(a_{p})-{\rm ev}_{0}^{*}(a_{p})+k(-1)^{p+1}
  \int_{0}^1 a_{p+1} \label{wedgep}\\
\genf b q &\stackrel{\rm def}{=}& I(b_q + b_{q+1}\zeta) =
   {\rm ev}_{1}^{*}(b_{q})- {\rm ev}_{0}^{*}(b_{q})+k(-1)^{ q+1}
  \int_{0}^1  b_{q+1} \label{wedgeq}
\end{eqnarray}
Then it follows that the wedge product on
  $I(\hat\Omega(X))$  is given by
\begin{eqnarray}
\label{exteriordeffinal}
\genf a p\wedge' \genf b q &=&  {\rm ev}_{1}^{*}(a_{p} b_{q}) -
 {\rm ev}_{0}^{*}(a_{p}b_{q}) \, \\
&& \nonumber \qquad +
k(-1)^{p+q+1}\int_{0}^1 \bigl(a_{p}b_{q+1}+(-1)^{q}a_{p+1}
b_{q}) \,.
\end{eqnarray}

Since $\wedge' $ is given through $I$ by means of the commuting
diagram  (\ref{CD:generalwedge1}), it follows that
\begin{eqnarray}
&& \label{leibniz} d (\genf a p\wedge' \genf b q)= (d  \genf a
  p)\wedge'   \genf b q+(-1)^p\genf a p\wedge  d \genf b  q\,,
\\
&&\label{commutation} \genf a p\wedge' \genf b q=(-1)^{pq} \genf b q
  \wedge' \genf a  p\,.
\end{eqnarray}
Note that $\wedge'$ is not simply the natural wedge product on
$\Omega(\mpx)$, though the two are presumably related.

\section*{Acknowledgements} ANS thanks the {\'E}cole Normale
Sup{\'e}rieure, Paris, and the SN Bose Center, Kolkata, for research
visits during which this work was done.

\appendix
\section{Appendix}\label{App}

We will summarize here  some essentials concerning graded algebras
and complexes, for ease of reference.

A graded vector space is a vector space $V$ expressed as a direct
sum of subspaces $V_k$, with $k$ running over the set $\mbz$ of
integers:
$$V=\bigoplus_{k\in{\mbz}}V_k$$ The non-zero elements of $V_k$ will
be said to have {\em degree} $k$, and we define the degree of $0$
arbitrarily to be $0$. Elements in some $V_k$ will be said to be
{\em homogeneous}, and the degree of a homogeneous element $x$ will
be denoted $d_x$ or $d(x)$.

 Often we are interested in the case of a {\em module} over a ring
$R$ (which could be the space of smooth functions on a manifold),
instead of a vector space. Here and henceforth $R$ is a commutative
ring with $1\neq 0$.  A {\em graded} $R$-algebra is a graded
$R$-module $$A=\bigoplus_{p\in{\mbz}}A_p$$ along with a bilinear,
associative product
$$A\times A\to A$$
for which
$$A_pA_q\subset A_{p+q}\qquad\hbox{for all $p,q\in\mbz$.}$$
The {\em super-commutator} in such a graded algebra is the bilinear
map
$$A\times A\to A: (a,b)\mapsto\{a,b\}$$
such that, for homogeneous $x$ and $y$,
\begin{equation}
\label{supercommutator}
 \{x,y\}=xy-(-1)^{d_xd_y}yx
\end{equation}
The algebra is {\em super-commutative} if all super-commutators are
zero. For instance, the smooth differential forms generate a
super-commutative graded algebra.

A {\em super-derivation} $D$ of degree $q\in\mbz$ on $A$ is an
$R$-linear map
$$D:A\to A: x\mapsto D(x)$$
for which
$D(A_j)\subset A_{j+q}$ for all $j\in\mbz$,
satisfying
\begin{equation}
\label{superderiv}
D(xy)=D(x)y+(-1)^{d_xq}xD(y)
\end{equation}
for all homogeneous elements $x,y\in A$.
If $D$ and $D'$ are superderivations of degrees $q$ and $q'$ then
$$ \{D,D'\}\stackrel{\rm def}{=} DD'-(-1)^{qq'}D'D$$
is a superderivation of degree $q+q'$.

Now consider   graded $R$-modules $A$ and $B$. There is then the
tensor product $R$-module
$$A\otimes B$$ whose elements are linear combinations of elements
of the form $a\otimes b$ with $a\in A$ and $b\in B$. Define a
${\mbz}$-grading on this tensor product by setting
$$(A\otimes B)_p=\bigoplus_{q\in{\mbz}}A_q\otimes B_{p-q}$$
If $A$ and $B$ are graded algebras, then $A\otimes B$ is also a
graded algebra under the product
\begin{equation}\label{gradedproduct}
(a\otimes b)(a'\otimes b)=(-1)^{d_bd_{a'}}(aa')\otimes
 (bb')\end{equation}
for all homogeneous $a,a'\in A$ and homogeneous $b,b'\in B$.  It
may be checked by a lengthy computation (or by functorial
arguments) that $A\otimes B$ is also associative.

It may be convenient to write $ab$ instead of $a\otimes b$. With
this convention, the product is defined through
$$aba'b'=(-1)^{d_bd_{a'}}aa'bb'$$
for homogeneous elements.

Now consider a superderivation $D$, of degree $q$, on $A$, and a
superderivation of the same degree on $B$, and, for notational
convenience, we denote the latter superderivation also by $D$.
Define a linear map, again denoted $D$,
$$D:A\otimes B\to A\otimes B$$
by requiring that
\begin{equation}
\label{defF}
D(ab)=D(a)b+(-1)^{qd_a}aD(b)
\end{equation}
for homogeneous $a\in A$ and $b\in B$.
Then, for homogeneous elements $a,b,a',b'$, we have
\begin{eqnarray*}
D(aba'b') &=& (-1)^{d(b)d(a')}D(aa'bb')\\
&=&(-1)^{d(b)d(a')}
\left[D(aa')bb'+(-1)^{q(d(a)+d(a'))}aa'D(bb')\right] \\
&=&(-1)^{d(b)d(a')} \left[D(a)a'bb'+(-1)^{qd_a}aD(a')bb'\right]\\
&&\quad + (-1)^{d(b)d(a')+q(d(a)+d(a'))}
\left[aa'D(b)b'+(-1)^{qd_b}aa'bD(b')\right]\\
&=& D(a) ba'b'+(-1)^{qd_a}aD(b)a'b'\\
&&\qquad
+(-1)^{qd(ab)}abD(a')b
 +(-1)^{qd(ab)+qd(a')}aba'D(b')\\
&=& D(ab)a'b'+(-1)^{qd(ab)}abD(a'b')
\end{eqnarray*}
Thus, $D$ is a superderivation of degree $q$ on $A\otimes B$.

Applying $D$ twice successively we have
\begin{eqnarray*}
D^2(ab)&=&D^2(a)b +(-1)^{q(q+d_a)}D(a)D(b)\\
&&\qquad +(-1)^{qd_a}D(a)D(b) +aD^2(b)
\end{eqnarray*}
Taking $q$ to be odd, we obtain: \par
 \noindent{\bf Proposition
A.1.} {\em If $D$ is a super-derivation of a common odd degree $q$
on each of the graded algebras $A$ and $B$, and $D^2=0$ on $A$ and
on $B$, then on $A\otimes B$ the induced $D$ is also a
super-derivation of degree $q$   satisfying $D^2=0$.}


\end{document}